\documentclass[aps,prl,reprint,amsmath,amssymb,groupedaddress,showpacs]{revtex4-1}
\usepackage{graphicx}
\usepackage{dcolumn}
\usepackage{bm}
\usepackage{hyperref}
\usepackage{amsmath,amssymb,amsfonts}
\usepackage{xcolor}
\usepackage{siunitx}
\usepackage{natbib}

\newcommand{\abs}[1]{\left|#1\right|}

\newcommand{\parti}[2]{\frac{\partial #1}{\partial #2}}
\newcommand{\partii}[2]{\frac{\partial^2 #1}{\partial #2^2}}

\begin{document}

\title{Observation of modulational instability in Bose-Einstein condensates}%

\author{P.J.~Everitt$^1$}
\author{M.A.~Sooriyabandara$^1$}
\author{M.~Guasoni$^{2}$}
\author{P.B.~Wigley$^1$}
\author{C.H.~Wei$^{1,3}$}
\author{G.D.~McDonald$^1$}
\author{K.S.~Hardman$^1$}
\author{P.~Manju$^1$}
\author{J.D.~Close$^1$}
\author{C.C.N.~Kuhn$^1$}
\author{S.S.~Szigeti$^4$}
\author{Y.S.~Kivshar$^2$}
\author{N.P.~Robins$^1$}

\affiliation{$^1$Department of Quantum Science, Research School of Physics and Engineering, Australian National University, Canberra ACT 2601, Australia\\
$^2$Nonlinear Physics Center, Research School of Physics and Engineering, Australian National University, Canberra ACT 2601, Australia\\
$^3$Department of Instrument Science and Technology, College of Mechatronic Engineering and Automation, National University of Defense Technology, Changsha 410073, China\\
$^4$Department of Physics, Centre for Quantum Science, and Dodd-Walls Centre for Photonic and Quantum Technologies, University of Otago, Dunedin 9010, New Zealand}

\begin{abstract}
We observe the breakup dynamics of an elongated cloud of condensed $^{85}$Rb atoms placed in an optical waveguide. The number of localized spatial components observed in the breakup is compared with the number of solitons predicted by a plane-wave stability analysis of the nonpolynomial nonlinear Schr\"odinger equation, an effective one-dimensional approximation of the Gross-Pitaevskii equation for cigar-shaped condensates. It is shown that the numbers predicted from the fastest growing sidebands are consistent with the experimental data, suggesting that modulational instability is the key underlying physical mechanism driving the breakup.
\end{abstract}

\maketitle

Intensity-dependent instabilities are a dramatic manifestation of the strong nonlinear effects that can occur in nature, and they are observed throughout physics as the development of spatial or temporal modulations with growing amplitudes. {\it Modulational instability} (MI) is a well-known phenomenon in optics which manifests itself as a decay of long optical signals into pulse trains~\cite{mi_optics,hasegawa_tunable_1980,agrawal_modulation_1987}. MI is a general wave phenomenon that occurs when a weak perturbation to a waveform is enhanced by nonlinearity, giving rise to sidebands in the spectrum with subsequent modulation growth and the formation of a train of spatially or temporally separated solitary waves~\cite{kivshar_optical_2003}.

Solitary matter-waves of a different origin have been extensively studied in Bose-Einstein condensates (BECs), where interatomic interactions give rise to strong nonlinearities. In particular, soliton trains in BECs have previously been observed in $^7$Li condensates with attractive interactions loaded into highly anisotropic traps~\cite{strecker_formation_2002}. The formation of multiple solitary waves has also been observed during collapse of Rb condensates~\cite{cornish_formation_2006}. Recently, BEC solitons have been employed for the first realization of a solitonic atom interferometer~\cite{mcdonald_bright_2014}. Despite theoretical predictions~\cite{Salasnich2003,Carr_2004}, the stochastic nature of many nonlinear processes, combined with traditionally destructive methods of BEC imaging, has impeded the direct observation of more subtle nonlinear effects.

\begin{figure}
    \includegraphics[width=\columnwidth]{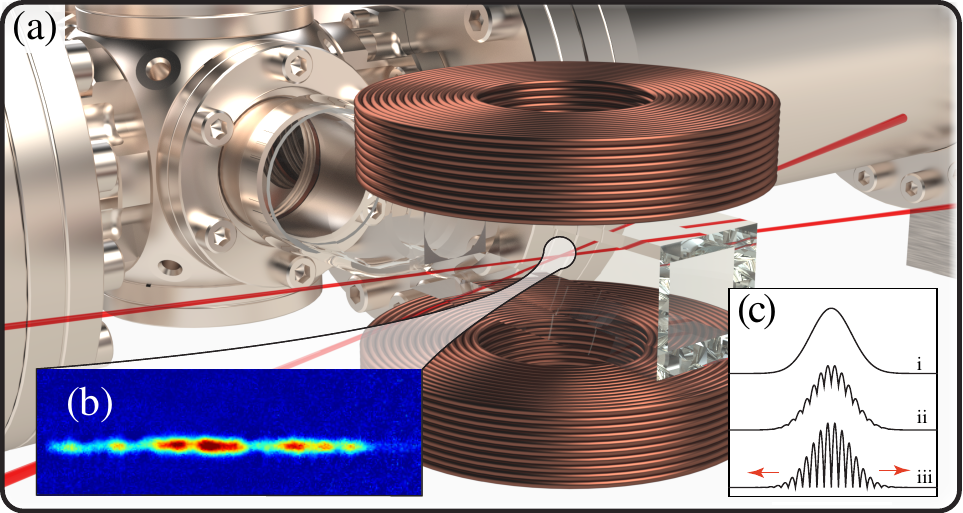}
    \caption{ \textbf{(a)} Schematic of UHV apparatus and an optical waveguide. \textbf{(b)} Soliton train observed after $\SI{100}{\milli\second}$ propagation of a BEC in the waveguide at $a_s=-1.15a_0$. \textbf{(c)} Modulational instability in a 1D model: (i) an initially localized state becomes (ii) modulated by unstable momentum sidebands which (iii) grow exponentially to produce a train of solitons.}
    \label{fig:expSetup}
\end{figure}

In this Rapid Communication, we present real-time observations of MI in BECs. Using nondestructive \emph{in situ} imaging, we are able to image a single BEC placed in an optical waveguide as it undergoes the transformation into a train of spatially localized components. We show that the number of localized components observed after the breakup is consistent with an MI analysis conducted in the framework of the nonpolynomial Schr\"odinger equation (NPSE)~\cite{Salasnich2002}, suggesting that MI is the underlying physical mechanism driving the observed breakup.

\begin{figure*}[t!]
 	\centering{}
 	\includegraphics[width=\textwidth]{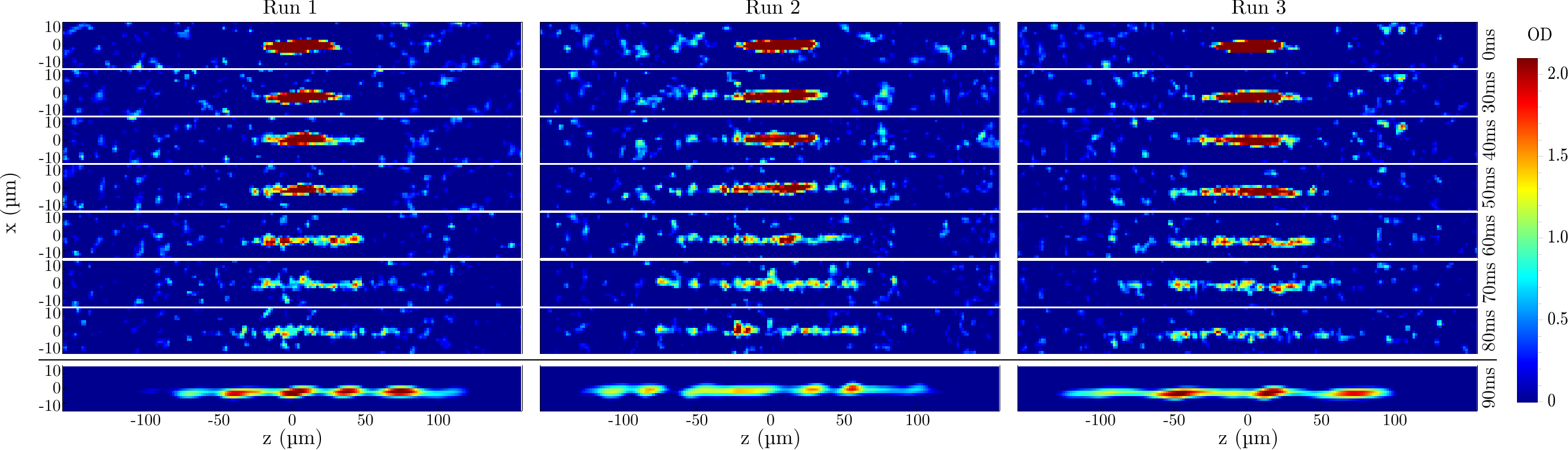}
 	\caption{Experimental data for three separate runs illustrating the breakup phenomenon after BEC is released into the waveguide with $a_s = -1.15a_0$. The first seven images were taken with \emph{in situ} nondestructive shadowgraph imaging. The final image was taken with \emph{in situ} destructive absorption imaging. Color map optical density (OD) values relate to the destructive picture.}
 	\label{fig:expBreakup}
 \end{figure*}

{\em Experimental approach.} The experimental apparatus is described in Ref.~\cite{kuhn_bose-condensed_2014}. In summary, a combined two- and three-dimensional MOT system collects and cools both $^{85}$Rb and $^{87}$Rb atoms (Fig.~\ref{fig:expSetup}). The atoms are then loaded into a magnetic trap and undergo RF evaporation before being loaded into an optical crossed-dipole trap. The cross trap consists of intersecting $\SI{1090}{\nano\meter}$ and $\SI{1064}{\nano\meter}$ laser beams with approximate waists of $\SI{300}{\micro\meter}$ and $\SI{250}{\micro\meter}$ (halfwidth at $1/e^2$ intensity), respectively. After loading, the magnetic trap coils are switched from antiHelmholtz to Helmholtz configuration, generating a bias magnetic field over the extent of the cloud. This allows the $s$-wave scattering length of the cloud, $a_s$, to be tuned utilizing a Feshbach resonance. Setting the $^{85}$Rb scattering length close to zero ($a_{s}=3a_{0}$, with $a_0$ the Bohr radius) while ramping down the cross trap intensity allows the remaining $^{87}$Rb atoms to be removed while minimizing three-body recombination losses in $^{85}$Rb. A further period of evaporation with $a_{s}=300a_{0}$ creates a $^{85}$Rb BEC of atom number $N=3\times 10^4$ with no observable thermal fraction. This is the initial condition for all experiments discussed below.

To monitor the cloud dynamics two orthogonal imaging systems are available. Firstly, a horizontal absorption beam allows the cloud to be imaged after 20ms of ballistic expansion ($a_{s}=0$) in free space to calibrate absolute atom number. A second vertical, far-detuned imaging beam utilizes nondestructive shadowgraph imaging to take \emph{in situ} images of the condensate~\cite{wigley_non-destructive_2016}. Up to 100 images can be taken in a single run as little as 0.4ms apart with no measurable change in atom number. The imaging laser is offset-beat locked, allowing the beam detuning to be dynamically changed during a run~\cite{wei_compact_2016}. This allows several nondestructive pictures to be taken before the laser is brought onto resonance and a final destructive, high signal-to-noise ratio (SNR) picture is taken with the same optics.

The experiment is conducted as follows: a $^{85}$Rb condensate is formed at an s-wave scattering length of $a_s\approx300 a_0$ in a harmonic trap ($\omega_z=2\pi\times7\;\mathrm{Hz}$ axially, $\omega_r=2\pi\times70\;\mathrm{Hz}$ radially) before the axial trapping is turned off creating an optical waveguide for the atoms. The waveguide has a repulsive harmonic potential with frequency $\omega_z=2\pi \times 3i\;\mathrm{Hz}$ due to curvature in the magnetic potential~\cite{mcdonald_bright_2014}. Simultaneously, the scattering length is rapidly quenched to another value ($a_{s}/a_{0}=[-2.1,5]$) before the condensate is allowed to evolve in the waveguide for approximately $100\si{\milli\second}$. In all figures $t$ is the time after this quench, once the condensate is freely propagating in the waveguide. For these initial conditions the chemical potential is $\mu = 6.2 \hbar \omega_z$, which is the strongly cigar-shaped regime~\cite{petrov_regimes_2000}. The longitudinal shape of the condensate has a width ($L_z$) on the order of tens of micrometers. Dispersion of the cloud in the waveguide is minimized by careful choice of initial scattering length. Our choice ensures that the dimensions of the axially-trapped cloud prior to quenching $a_s$ are similar to those of an axially-untrapped soliton at the new (necessarily negative) scattering length. This is determined by a variational method~\cite{Carr2002}. The condensates are then propagated in the waveguide for a range of scattering lengths ($a_s/a_0\in[-6,6]$).

For propagation at certain negative scattering lengths the condensate breaks up into a train of similarly sized, spatially-localized components. These are soliton-like, in that they are stable under further propagation beyond $90\si{\milli\second}$. Three separate exemplar runs  are shown in Fig.~\ref{fig:expBreakup}. These examples demonstrate the typical train formation at $-1.15a_{0}$, a scattering length for which the condensate is stable up to ~60ms. Moreover, for $a_s>-1.15a_{0}$, three-body-recombination losses are $<5 \%$ in the first-stage dynamics where train formation occurs.

Ten runs were conducted for each choice of $a_s$. The number of spatial components was quantified using an image processing algorithm (see Supplemental Material \cite{supplemental}). Briefly, the images were de-noised then binarized with the number of resulting solitons given by counting the morphological components. Train formation is stochastic in nature -- the same experimental conditions resulted in a varying number of constituent components (Fig.~\ref{fig:expBreakup}).  Although the atom number varied shot-to-shot by up to 10\%, this variation was shown to be uncorrelated with the final number of components.  Variations in both the spatial locations of the individual components and the formation onset time were also observed.

{\em Theoretical approach.} The NPSE~\cite{Salasnich2002} -- an effective 1D model of the Gross-Pitaevskii equation (GPE) -- provides a simple and effective theoretical insight into our experimental observations. It well-approximates the full 3D dynamics of the GPE for cigar-shaped condensates whose relevant dynamics occur in the axial direction~\cite{Salasnich2002,Salasnich2003}. According to the NPSE, the 3D condensate wavefunction is factorized into transverse and longitudinal components: $\psi(x,y,z,t) = f(z,t)\phi(x,y,t; f(z,t))$.
The axial component $f(z,t)$ is the unique unknown of the model, and its evolution is described by
\begin{equation}
i\parti{f}{t} = \left[ -c_1\frac{1}{2}\partii{}{z}  + \frac{c_2}{\sigma^2}\abs{f}^2 + c_3\left(\sigma^2 + \frac{1}{\sigma^2}\right) \right]f,
\label{eq:nlse_f}
\end{equation}
where $c_1=\hbar/(2m)$, $c_2=U N/(2\pi a_\perp^2 \hbar)$, $c_3=\omega_\perp/2$ and  $\sigma^2=(1+c_4\abs{f}^2)^{1/2}$, with $c_4=2a_sN$. Here $U=4\pi \hbar^2 a_s/m$, $N$ is the condensate number, and $a_\perp= \sqrt{\hbar / (m\omega_\perp)}$ is the oscillator length in the transverse direction with corresponding trapping frequency $\omega_\perp$. 

Following the standard procedure \cite{hasegawa_tunable_1980}, we describe the MI dynamics in the axial coordinate by assuming $f(z,t)$ is the sum of a plane wave and two small perturbations centered at spatial frequencies $\pm k$: $f(z,t) = u \exp(i \beta_u t) + p_+ \exp [i k z + i \beta_+(k) t + g(k) t] + p_- \exp [-i k z + i \beta_-(k) t + g(k) t]$, where $u,p_\pm$ are complex, $\beta_u$ is real, and $\beta_\pm$ and $g$ are real functions of $k$, with $g$ the temporal growth rate of the perturbations.

Inserting this ansatz into Eq.~\eqref{eq:nlse_f} and applying a linear stability analysis yields expressions for $\beta_{u,\pm}$ and $g$. The latter quantity is the gain spectrum of spatial modes with wavenumber $k$:
\begin{align}
	g(k) 	&= \sqrt{c_1 k^2 (4M-c_1 k^2)} / 2, \\
	M		&=\frac{-(c_2+c_3c_4/2)\abs{u}^2}{\sigma_0^2}+\frac{(c_2\abs{u}^2+c_3)c_4\abs{u}^2}{2\sigma_0^6},
\label{eq:nlse_gain}
\end{align}
where  $\sigma_0^2=(1+c_4\abs{u}^2)^{1/2}$. 

Equation~\eqref{eq:nlse_gain} implies that all spatial frequencies in the MI-band $|k| \leq 2(M/c_1)^{1/2}$ undergo amplification and that the fastest growing sidebands are at $\hat{k}=(2M/c_1)^{1/2}$ with corresponding growth rate $\hat{g}=M$.

In the weakly-interacting limit $c_4\abs{f}^2\ll 1$, Eq.~\eqref{eq:nlse_f} reduces to the well-known 1D GPE \cite{Salasnich2002}, which is less accurate than the NPSE as it fails to describe the condensate dynamics outside the weakly-interacting regime~\cite{Salasnich2002}. Therefore, Eq.~\eqref{eq:nlse_gain} accurately describes the growth rate of perturbations in both weak- and strongly-interacting regimes.
\begin{figure}[t!]
	\includegraphics[width=\columnwidth]{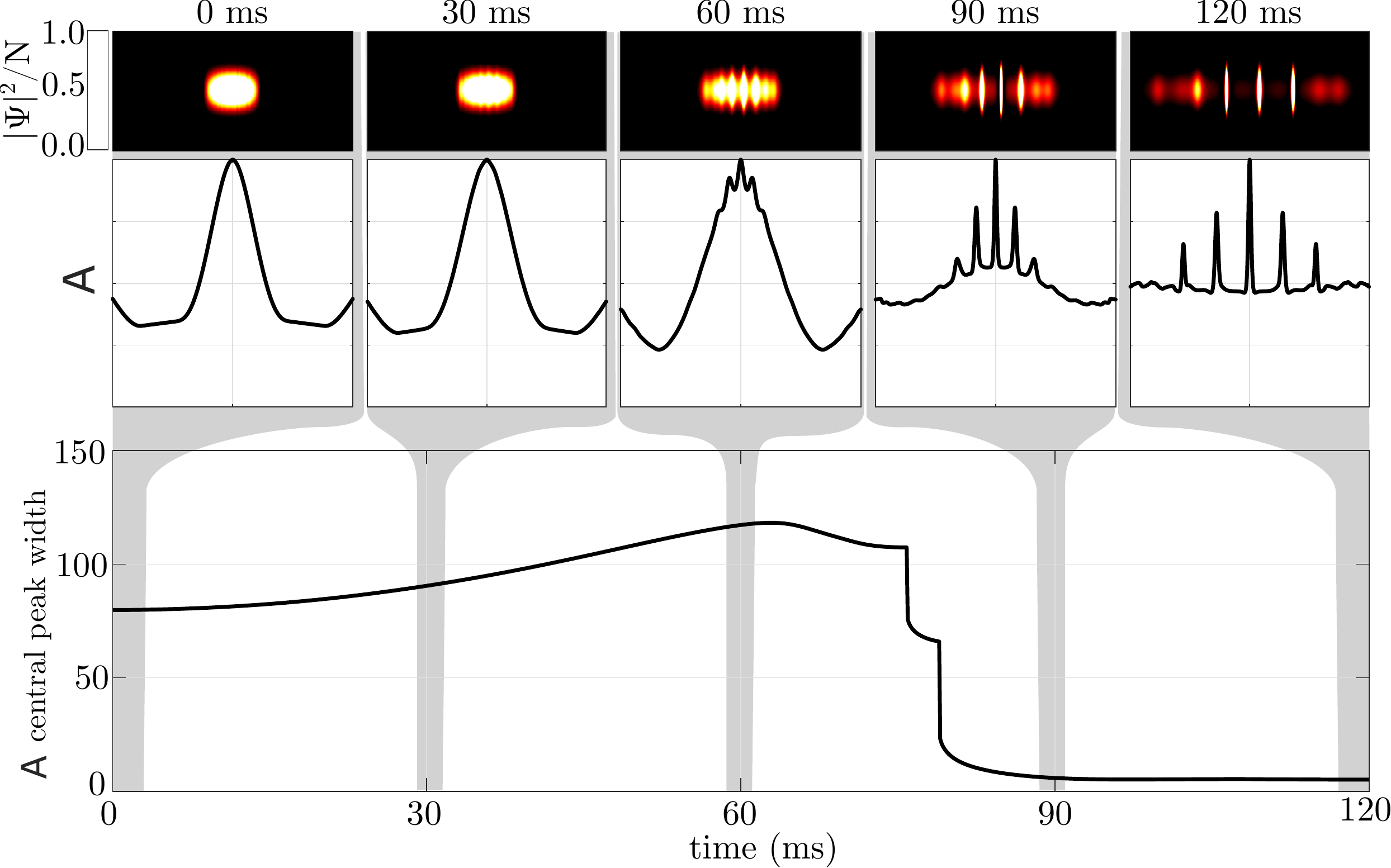}
	\caption{
		\label{fig:numerics}
		\textit{Top:} Density of a simulated BEC undergoing MI in an antitrapping waveguide. 
\textit{Middle:} Autocorrelation traces of axial density: $\mathcal{A}(z) = \int dz'|f(z-z')|^2 |f(z')|^2$. \textit{Bottom:} Width of central peak of $\mathcal{A}(z)$. This width's sharp decrease at $\SI{75}{\milli\second}$ corresponds to the time where MI is well-developed.}
\end{figure}

In the more general and realistic case where the initial axial function $u_0(z)=f(z,t=0)$ is not a plane wave but finite-sized, the BEC undergoes both a linear reshaping [first term in RHS of Eq.~\eqref{eq:nlse_f}] and a nonlinear reshaping [second and third terms in RHS of Eq.~\eqref{eq:nlse_f}], which are respectively analogous to the diffraction and nonlinear self-phase modulation (SPM) experienced by an optical beam propagating in nonlinear Kerr-media~\cite{Dudley2006}. 

Linear (diffraction) and nonlinear (SPM and MI) effects interplay with each other and a characteristic time defines the temporal scale over which they become relevant to the dynamics \cite{supplemental}. Exploiting the analogy with optical phenomena in Kerr media, the characteristic time for linear diffraction and nonlinear phenomena can be roughly estimated as $T_\textrm{D} = L_z/c_1^2 $ and $T_\textrm{NL}=1/(\abs{c_2}u_\text{peak}^2)$, respectively, where $L_z$ is the spatial width of $u_0$ and $u_\text{peak}=\max\{\abs{u_0}\}$. In our experiment $T_\textrm{D} \gg T_\text{NL}$, hence the first-stage evolution of the condensate is dominated by MI and can be written as:
\begin{eqnarray}
f(z,t) = u_0e^{i\beta_u t} + \int dk \, p(k) e^{ikz+i\beta(k) t +g(k) t}
\label{eq:MI_evolution}
\end{eqnarray}
where the integration is performed over the whole MI-band and $p(k)$ indicates the initial perturbation at spatial frequency $k$ that is amplified with growth rate $g(k)$ [see Eq.~\eqref{eq:nlse_gain}]. The integral in Eq.~\eqref{eq:MI_evolution} represents a superposition of sinusoidal waves giving rise to the typical components observed in the condensate axial profile. If the condensate is seeded by random noise, then the initial perturbation $p(k)$ is a stochastic variable whose value depends on the particular experimental run. Consequently, for a fixed time $t$, the position, amplitude, and shape of the components differs between runs. This explains the random nature of the components experimentally observed and represents a truly distinctive signature of the MI regime.

Although the number of morphological components after breakup is random, the average distance $\langle d_z \rangle$ between consecutive components is set by the fastest growing spatial frequency, i.e. $\langle d_z \rangle = 2 \pi /\hat{k}$. This allows a simple estimate of the average number of components via $\langle N_\text{Comp} \rangle=L_z/ \langle d_z \rangle$.  

\begin{figure}[!t]
	\centering{}
	\includegraphics[width=\columnwidth]{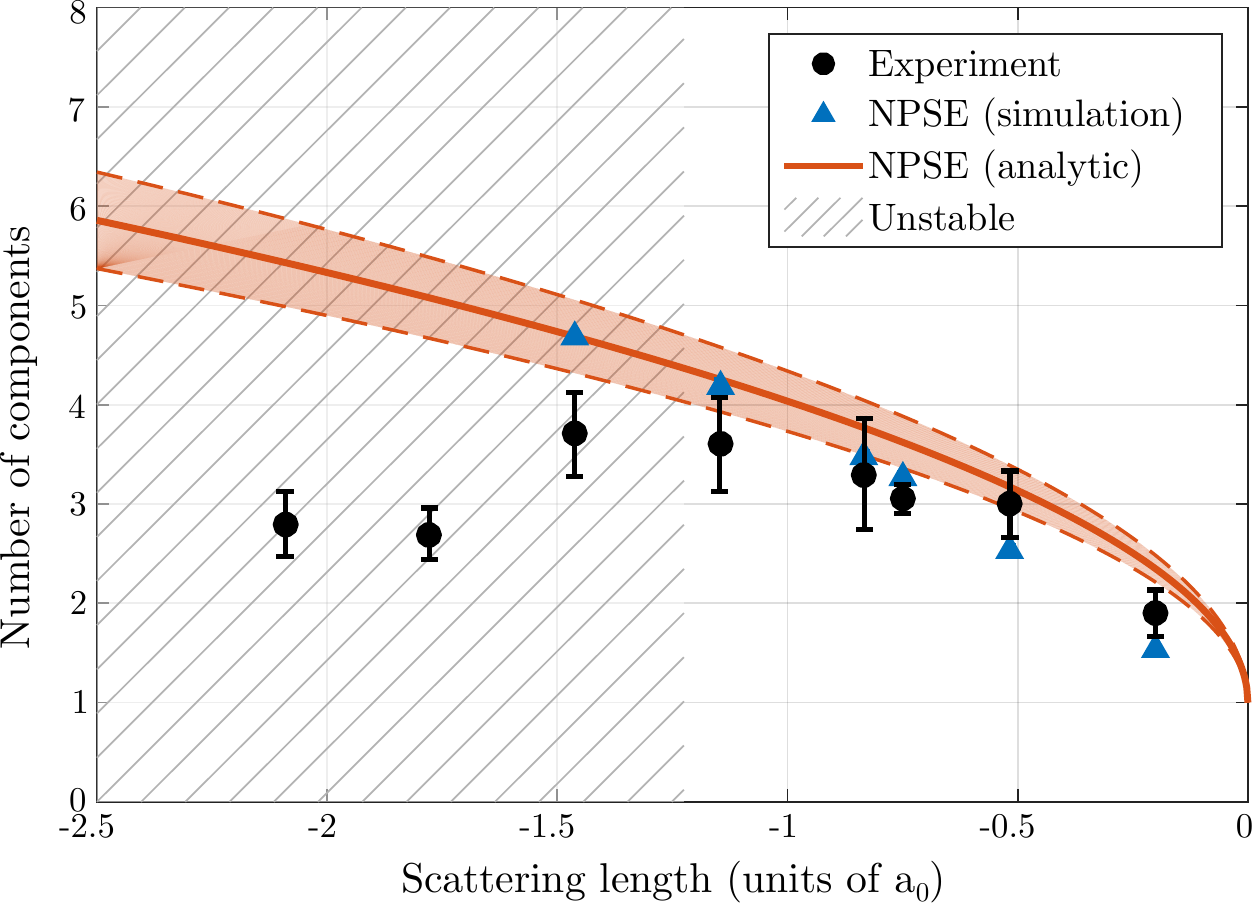}
	\caption{Average number of spatial components observed in experiment at t=$90 $ms compared with NPSE simulations and analytic estimation $\langle N_\text{Comp} \rangle= L_z/\langle d_z \rangle$ (shading indicates 10\% uncertainty in measured $L_z$). The initial BEC is considered to have one component. Uncertainties are indicated by the standard error in the mean (these are smaller than the point width for simulations). The hatched region indicates the absence of a stable soliton solution.}
	\label{fig:npseSidebands}
\end{figure}

Additional to this analytic theory, we performed NPSE simulations that aimed to reproduce experimental conditions, including the antitrapping waveguide potential and three-body recombination losses, modeled by including $iK_3 N^2 |f|^4 f/(2\pi a_\perp^2\sigma^2)^2$ in the RHS of Eq.~\eqref{eq:nlse_f} ($K_3=4\times10^{-41}\mathrm{\;m}^6\mathrm{s}^{-1}$ is experimentally determined). The simulations were seeded with Gaussian noise with a minimum spatial correlation width determined by the condensate's healing length, ensuring that the perturbation only contained physical noise. The noise amplitude was approximately 1\% of $|\psi|$.

Figure~\ref{fig:numerics} displays the evolution of a simulated BEC undergoing MI, resembling the typical evolution observed in experiments (Fig.~\ref{fig:expBreakup}). Condensate breakup causes the formation of spatial components, which become distinct and separated once MI amplification is well-developed. Furthermore, the final distribution is typically asymmetric as a result of the random seeding noise ($|\Psi|^2$ in Fig.~\ref{fig:numerics}).

SPM induces an axial compression of the condensate, corresponding to a broadening of the condensate's initial axial spectrum (i.e. Fourier transform of $u_0$). The larger the width $\Delta k$ of the initial axial spectrum, the faster the spectral broadening and the corresponding spatial compression. When $\Delta k \gtrsim \hat{k}$, the spectral broadening of the axial spectrum overlaps the MI-band and amplification of the seeding noise is negligible. Here, the SPM-induced spatial compression is followed by complex dynamics that could lead to the creation of distinct spatial components \cite{supplemental}. These dynamics are similar to those undergone by short optical pulses in Kerr-media~\cite{Dudley2006}. However, in this circumstance the number, position, and shape of the spatial components are fully deterministic, as they do not depend on the seeding noise.

An intermediate regime exists where neither MI nor SPM are truly dominant, but are equally important to the condensate dynamics. In this instance, the condensate undergoes a partial reshaping followed by the creation of random spatial components related to the MI \cite{supplemental}. 

For typical experimental parameters $L_z\sim 50 \mu$m and $a_s \sim- a_0$, we estimate $\Delta k \sim 1/L_z \approx 20$m$^{-1}$ and $\hat{k}\approx 100 \Delta k$, therefore SPM does not dominate over MI. However, for sufficiently negative scattering lengths, simulations show that a fast and strong spectral broadening due to SPM occurs.

\begin{figure}[t!]
	\centering{}
	\includegraphics[width=0.9 \columnwidth]{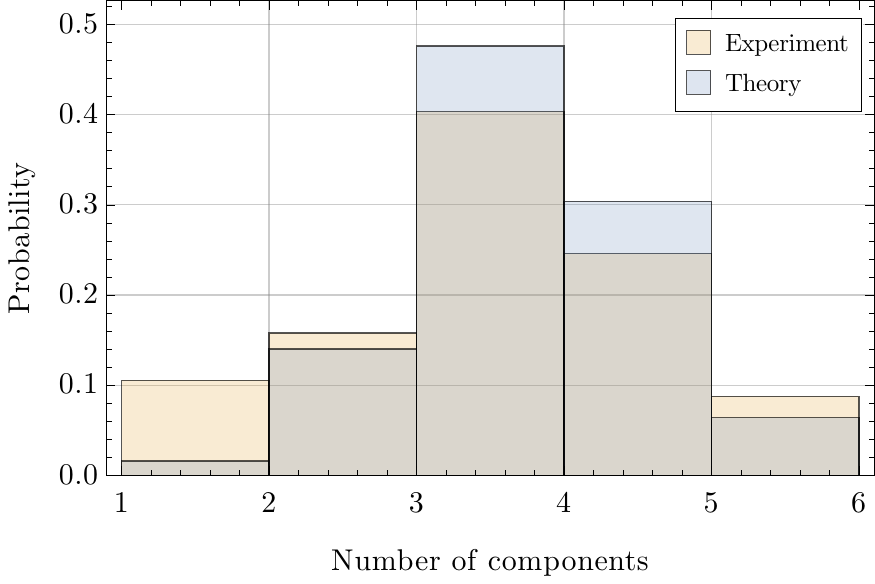}
	\caption{Probability of observing a given number of spatial components for both experiment (57 runs) and simulations (250 runs) at quenched scattering length $a_s = -0.75a_{0}$. Grey regions indicate overlap of both distributions. An independent sample $t$-test ($t(71.67) = -1.346$, $p=.183$) reveals no significant difference between the two distributions. 
    }
	\label{fig:histogram}
\end{figure}

The average number of spatial components experimentally observed as a function of quenched $a_s$ is shown in Fig.~\ref{fig:npseSidebands} alongside the predictions from a stability analysis of the NPSE as well as NPSE simulations. The difference between simulations and analytic estimation is expected due to the presence of a partial SPM pulse reshaping as well as the antitrapping potential not accounted for by our analytical model. Nevertheless, after excluding data for $a_s < -1.7 a_0$ (where a large thermal population was observed, so mean-field formalism is not applicable), $\chi^2$-tests show that our analytic theory ($\chi^2(6) = 1.512$, $p = .959$) and NPSE simulations ($\chi^2(6) = 0.861$, $p = .990$) agree well with the experimental data. Note also that stable soliton solutions of the initial condensate do not exist below $a_s=-1.2 a_0$, as determined by a variational analysis of the GPE energy functional~\cite{Carr2002}.

The \emph{distribution} of measured components was also compared to that predicted by NPSE simulations. Independent sample $t$-tests show that the distribution of component numbers predicted by the NPSE do not significantly differ to the experimentally-measured distributions \cite{supplemental}; this agreement is pictorially demonstrated for $a_s = -0.75 a_0$ in Fig.~\ref{fig:histogram}.

{\em Discussion.} The observed random position, number, and amplitude of measured components is truly indicative of MI dynamics. Agreement with NPSE simulations of MI is seen in both the mean and standard deviation of the number of measured components. In contrast, if train formation was seeded by linear effects, its dynamics would be deterministic, similar to what occurs with SPM reshaping. Furthermore, although the mean-field formalism employed here cannot account for fragmentation effects captured by the many-body model, the fragmentation dynamics predicted in an initially symmetric condensate are highly symmetric~\cite{Cederbaum08}. Therefore, the stochastic breakup dynamics observed in our experiments are a truly distinctive signature of MI, whereas the observation of symmetric and deterministic behavior would be suggestive of a different underlying mechanism, such as MI seeded by linear effects from fragmentation or SPM reshaping.

In summary, we have presented a continuous experimental observation of MI in a BEC. A major advance of our experiment is the ability to nondestructively image stochastic time-dependent nonlinear phenomena with high temporal resolution. Our experimental observations are in good agreement with analytical and numerical predictions provided by the NPSE, and suggest that MI seeded by noise is the key physical mechanism underlying the breakup of elongated condensates into matter-wave solitary waves.

The improvements made to the experimental apparatus will allow the detailed study of many new aspects of nonlinear dynamics driven by MI. For MI to occur, an initial perturbation is required that then exponentially grows. Applying a reproducible initial noise profile to seed the condensate (necessarily greater in amplitude than those already present in the system) would remove the stochastic nature of the MI, allowing the timescales and transient dynamics of MI to be closely compared with simulation. Additionally, the solitons here are created by matching the dimensions of the trapped and untrapped clouds when switching off the confining potential and adjusting the scattering length. If, instead, a small mismatch in the radial widths of these two solutions is created, then a transverse excitation will be present during MI, altering the dynamics in a way not captured by the NPSE. Finally, the role of relative phase in soliton-soliton interactions, either attractive or repulsive, has previously only been inferred through comparison with GPE simulation~\cite{nguyen_collisions_2014}. Combining the agile imaging system of the apparatus and the ability to perform Bragg interferometry along the waveguide~\cite{mcdonald_bright_2014} would allow a direct measurement of this effect. By first nondestructively observing train formation and subsequent soliton interactions, neighboring solitons can then be interfered and their relative phase read out with higher SNR destructive imaging.

Similar work, developed independently, was released after the preparation of this manuscript, which focuses on MI in Li condensates~\cite{nguyen_formation_2017}.

\section{Acknowledgements}

We acknowledge fruitful discussions with Tegan Cruwys and Michael Hush. This work was supported by the Australian Research Council and NanoPhi program of the European Union. S.S.S. received funding from an Australia Awards-Endeavour Research Fellowship and the Dodd-Walls Centre for Photonic and Quantum Technologies.


\begin{thebibliography}{99}

\bibitem{mi_optics} V.~I. Bespalov and V.I. Talanov, Filamentation structure of light beams in nonlinear liquids,
JETP Lett. {\bf 3}, 307 (1966).

\bibitem{hasegawa_tunable_1980}
A. Hasegawa and W.~F. Brinkman, Tunable coherent IR and FIR sources utilizing modulational instability,
IEEE J. Quantum Electron. {\bf 16}, 694 (1980).

\bibitem{agrawal_modulation_1987}
G.~P. Agrawal, Modulational instability induced by cross-phase modulation,
Phys. Rev. Lett. {\bf 59}, 880 (1987).

\bibitem{kivshar_optical_2003}  Y.S. Kivshar and G. Agrawal, {\em Optical Solitons: From Fibers to
Photonic Crystals} (Academic Press, Boston, 2003).

\bibitem{strecker_formation_2002} K.~E. Strecker, G.~B. Partridge, A.~G. Truscott, and R. Hulet, Formation and propagation
of matter-wave soliton trains, Nature {\bf 417}, 150 (2002).

\bibitem{cornish_formation_2006}  L. Cornish, S.T. Thompson, and C.E. Weiman, Formation of bright matter-wave solitons
during the collapse of attractive Bose-Einstein condensates, Phys. Rev. Lett. {\bf 96}, 170401 (2006).

\bibitem{mcdonald_bright_2014}  G.D. McDonald, C.C.N. Kuhn, K.S. Hardman, S. Bennetts, P.J. Everitt,
P.A. Altin, J.E. Debs, J.D. Close, and N.P. Robins, Bright solitonic matter-wave interferrometer,
Phys. Rev. Lett. {\bf 113}, 013002 (2014).

\bibitem{Salasnich2003}  L. Salasnich, A. Parola, and L. Reatto, Modulational instability and complex dynamics of
confined matter-wave solitons, Phys. Rev. Lett. {\bf 91}, 080405 (2003).

\bibitem{Carr_2004} L.D. Carr and J. Brand, Spontaneous soliton formation and modulational instability in Bose-Einstein condensates,
Phys. Rev. Lett. {\bf 92}, 040401 (2004).


\bibitem{Salasnich2002} L. Salasnich, A. Parola, and L. Realto, Effective wave equations for the dynamics of
cigar-shaped and disk-shaped Bose condensates, Phys. Rev. A {\bf 65}, 043614 (2002).

\bibitem{kuhn_bose-condensed_2014} C.C.N. Kuhn, G.D. McDonald, K.S. Hardman, S. Bennetts, P.J. Everitt,
P.A. Altin, J.E. Debs, J.D. Close, and N.P. Robins, A Bose-condensed, simultaneous dual-species Mach-Zehnder
atom interferometer, New J. Phys. {\bf 16}, 073035 (2014).

\bibitem{wigley_non-destructive_2016}  P. B. Wigley, P. J. Everitt, K. S. Hardman,  M. R. Hush, C. H. Wei, M. A. Sooriyabandara, P. Manju, J. D. Close, N. P. Robins, and C. C. N. Kuhn, Non-destructive shadowgraph imaging of ultra-cold atoms, Optics Letters {\bf 41}, 4795-4798 (2016).

\bibitem{wei_compact_2016} C. Wei, S. Yan, A, Jia, Y. Luo, Q, Hu, and Z. Li, Compact phase-lock loop for external
cavity diode lasers, Chinese Opt. Lett. {\bf 14}, 051403 (2016).

\bibitem{petrov_regimes_2000} D. S. Petrov, G. V. Shlyapnikov, and J. T. M. Walraven, Regimes of Quantum Degeneracy in Trapped 1D Gases, Phys. Rev. Lett {\bf 85}, 3745 (2000)

\bibitem{Carr2002} L.D. Carr and Y. Castin, Dynamics of a matter-wave bright soliton in
an expulsive potential, Phys. Rev. A {\bf 66}, 063602 (2002).



\bibitem{supplemental} See supplemental material for details on our soliton-counting algorithm, the interplay of the MI and reshaping regimes, and a summary of our $t$-tests.

\bibitem{Dudley2006} J.M. Dudley, G. Genty, and S. Coen, Supercontinuum generation in photonic crystal fiber, Rev. of Mod. Phys. {\bf 78}, 1135 (2006)
 
\bibitem{Cederbaum08} L.S. Cederbaum, A.I. Streltsov and O.E. Alon, Fragmented Metastable States Exist in an Attractive Bose-Einstein Condensate for Atom Numbers Well above the Critical Number of the Gross-Pitaevskii Theory, Phys. Rev. Lett {\bf 100}, 040402 (2008)
 
\bibitem{nguyen_collisions_2014} J.H.V. Nguyen, P. Dyke, De Luo, B.A. Malomed, and R.G. Hulet,
Collision of matter-wave solitons, Nature Physics {\bf 10}, 918-922 (2014).

\bibitem{nguyen_formation_2017} J.H.V. Nguyen, De Luo, and R.G. Hulet,
Formation of matter-wave soliton trains by modulational instability,  Science {\bf 356}, 422 (2017).

\end{thebibliography}
\end{document}